# Sheaf-theoretic representation of the proteolipid code

Troy A. Kervin[1,2,3]

[1] Division of Structural Biology, Nuffield Department of Medicine, University of Oxford, Oxford, OX3 7BN, UK
[2] Correspondence: troy.kervin@magd.ox.ac.uk
[3] ORCID: 0000-0002-7681-6861

## Abstract

Membrane particles such as proteins and lipids organize into zones that perform unique functions. Here, I introduce a topological and category-theoretic framework to represent particle and zone intra-scale interactions and inter-scale coupling. This involves carefully demarcating between different presheaf- or sheaf-assigned data levels to preserve functorial structure and account for particle and zone generalized poses. The framework can accommodate Hamiltonian mechanics, enabling dynamical modeling. This amounts to a versatile mathematical formalism for membrane structure and multiscale coupling.

## Introduction
The proteolipid code was recently proposed as a unifying model of membrane structure and function, countering the view that no such framework exists[1-3]. It describes membranes as being partitioned into zones which originate from interactions between proteins, lipids, and other membrane components. Each protein together with its lipid "fingerprint" represents one type of zone[4]. When proteins cluster, they create "protein island" zones enriched in lipids belonging to the fingerprints of their constituent proteins[5]. Zones that are devoid of protein are called "voids", which expand when proteins cluster[6]. Each zone has a unique structure and is coupled to other zones. For example, the composition of a lipid fingerprint depends on the surrounding void zone. The proteolipid code also offers a level of description sufficient to challenge previously proposed models as well as original and modified lipid raft theories[1,7-9].

Here, I advance the proteolipid code using category theory, which is concerned with relationships between objects[10,11]. Specifically, I employ sheaf theory[12,13] to assign data to open sets on a topological space, representing membrane anatomy at the particle and zone scales. Sheaf theory is introduced because it provides a principled way to assign and relate properties on overlapping regions while enforcing consistency between local and global structure, allowing zones to be coherently assembled from interacting particles. While sheaf theory has not, to the best of my knowledge, been used to model biological membrane structure itself, related topological methods have been used to characterize membrane structure for machine learning[14]. Existing formulations, though powerful for data analysis[15,16], require adaptation for complex physical systems. To address this, I extend traditional sheaf formulations to respond to particle- and zone-wise "poses" that influence interactions. My approach is compatible with learnable restriction maps in applied topology[17] and shares intuition with factorization algebras[18,19].

When addressing complex systems like membranes, the discussion often turns to emergent properties; a concept that remains philosophically contentious[20]. This debate appears to be partially fueled by canonical definitions of emergent properties being too broad[21], which can be counterproductive to delineating property subtypes. Rather than relying on emergence as an explicit concept, I formalize a hierarchical system of attributes proposed previously[1], generalizing the primary–secondary–tertiary–quaternary classification familiar from molecular biology. Importantly, this hierarchy is not intended as a mere relabeling, but as an operational decomposition in which properties are assigned to different supports. Unlike other generalizations[22], this hierarchy is centered on zones and their interactions, as represented mathematically with hypergraphs and category theory.

## Topology

Let $X \subset \mathbb{R}^d$ be a topological space representing the membrane. Define structured open set triples

$$U_i^* = (U_i, Y_i, T_i), \tag{1}$$

to cover $X$, where $U_i \subset X$ is an open set generated by a graded membership function, $Y_i$ is pose data, and $T_i$ is type. The graded membership functions that generate $U_i$ are stored in $Y_i$. Specifically, each $U_i$ is defined as a superlevel set of a membership function $\theta_i : X \to [0,1]$ at a fixed threshold[23]. We assume the membership functions $\theta_i$ are continuous (or at least lower semicontinuous), which ensures the resulting superlevel sets are open in $X$. The pose $Y_i$ therefore contains information necessary to generate the topology as well as additional geometric or physical measures such as orientation, which can be defined using a standard system such as normal vectors or quaternions[24,25]. I therefore use "pose" in a generalized sense and avoid exhaustively listing its elements since they are flexible. The type $T_i$ should specify the sort of object, such as 34:6 monogalactosyldiacylglycerol for a particle or caveola for a zone. $T_i$ can also be used as a label specifying the data that a sheaf assigns, while $Y_i$ will play a role of connecting sheaf-assigned and topological data to help determine how objects interact. Intersections can also have poses and types inherited from their constituent $U_i^*$s. For an index set $I = \{i_0, \dots, i_k\}$:

$$U_I = U_{i_0} \cap \dots \cap U_{i_k} \neq \emptyset, \tag{2}$$
$$Y_I = Y_{comp}(Y_{i_0} \dots Y_{i_k}), \tag{3}$$
$$T_I = T_{comp}(T_{i_0} \dots T_{i_k}), \tag{4}$$
$$U_I^* = (U_I, Y_I, T_I). \tag{5}$$

The overlap type $T_I$ may, for example, specify an interaction class, while the overlap pose $Y_I$ may encode analogous contextual information to that stored in the individual $Y_i$. The maps $Y_{\text{comp}}$ and $T_{\text{comp}}$ are included solely to indicate where application-specific overlap labeling may be introduced if needed; no aspect of the downstream sheaf-theoretic construction or restriction maps depends on their algebraic properties, and they may be omitted entirely in contexts where overlap pose or type labels are unnecessary. In applications where consistent overlap labeling is required across multiple intersections, one may impose additional conditions on $Y_{\text{comp}}$ and $T_{\text{comp}}$, such as associativity or commutativity.

Each $U_i^*$ and $U_I^*$ are physically meaningful regions that encode interaction capabilities. However, they are lacking in data, which will later be assigned by presheaves or sheaves. One motivation for treating poses and types as elements of structured open sets, while assigning attributes through a presheaf or sheaf, is that the latter provides a controlled mechanism for restricting and, when appropriate, gluing data across overlapping regions[12]. By contrast, pose and type belong to the underlying structural description of the system and are not required to satisfy gluing conditions, as they parameterize interaction context rather than represent quantities that must be globally consistent.

Define two structured open covers to represent particles (P) and zones (Z), respectively:

$$\mathcal{U}^P \equiv \{U_i^* \mid T_i \in \mathcal{T}^P\}, \tag{6}$$
$$\mathcal{U}^Z \equiv \{U_i^* \mid T_i \in \mathcal{T}^Z\}, \tag{7}$$

where $\mathcal{T}^P$ and $\mathcal{T}^Z$ are particle and zone supertypes, respectively. The zone cover should be constructed such that for every particle structured open set $U_a^* \in \mathcal{U}^P$ there exists a zone structured open set $U_\alpha^* \in \mathcal{U}^Z$ such that $U_a \subseteq U_\alpha$, ensuring that the particle cover is a refinement of the zone cover. To ensure that $\mathcal{U}^P$ and $\mathcal{U}^Z$ cover all areas of interest, they should also include "holes" in the membrane, which may implicitly refer to solvent or environment objects. Thus, on the particle scale, allow four main types: 1) Proteins, 2) Lipids, 3) Other particles, and 4) Holes. Likewise, the main types on the zone scale are: 1) Protein singletons with their associated lipid fingerprints, which will be called Units, 2) Protein islands and their collective lipid fingerprints, which will be called Isles, 3) Voids with lipids but no protein, and 4) Holes.

$$\mathcal{T}^P = \{\text{Protein, Lipid, Other, Hole}\}, \tag{8}$$
$$\mathcal{T}^Z = \{\text{Unit, Isle, Void, Hole}\}. \tag{9}$$

Throughout, indices $i$ and $j$ will denote general objects. Other lowercase Latin letters ($a, b,...$) will denote particles, and Greek letters ($\alpha, \beta,...$) will denote zones. Membership of points in particle and zones is defined via thresholding of the corresponding membership functions $\theta_i$:

$$x \in U_a \Leftrightarrow \theta_a(x) > \tau^P, \tag{10}$$
$$x \in U_\alpha \Leftrightarrow \theta_\alpha(x) > \tau^Z, \tag{11}$$

where $\tau^P, \tau^Z \in (0,1)$ are the thresholds for particle and zone membership, respectively. On the zone scale, the membership function $\theta_\alpha$ may be constructed using separate functions $\mu_a$ or $\nu_a$ stored in $Y_\alpha$ that determine the extent to which points $x$ belong to the fingerprint of protein $a$ or the void zone, respectively. I will provide one viable construction of membrane zonation under the assumption that Units and Isles do not extend beyond their fingerprints (Figure 1). In the following, assume $T_a \in$ Protein. Denote the threshold for belonging to fingerprints of Unit and Isle zones as $\tau^{\text{print}}$:

$$\forall a, \exists! \alpha \text{ such that } T_\alpha \in \{\text{Unit, Isle}\} \text{ and } \{x \mid \mu_a(x) > \tau^{\text{print}}\} \subseteq U_\alpha \tag{12}$$

Uniqueness in (12) is per protein: fingerprints of distinct proteins may partially overlap without inducing a new zone. An Isle zone is declared only when two or more proteins interact sufficiently through their fingerprints as quantified by a threshold $\tau^{\text{isle}} \geq \tau^{\text{print}}$:

$$\exists n \geq 1, \exists \text{distinct } a_0, ..., a_n \ (\forall i \in \{0, ..., n-1\}, \exists x_i \text{ such that } \mu_{a_i}(x_i) > \tau^{\text{isle}} \\ \text{and } \mu_{a_{i+1}}(x_i) > \tau^{\text{isle}}) \Rightarrow \exists! \alpha \text{ with } T_\alpha \in \text{Isle}. \tag{13}$$

Isles are formed for proteins connected under the relation "co-exceeding $\tau^{\text{isle}}$ at some $x$"; an Isle contains all proteins in the cluster. The points $x_i$ may differ for different pairs; a chain of pairwise co-exceedances (possibly at different locations) places all $a_i$ in one connected Isle. Finally, denote the threshold for void association by $\tau^{\text{void}}$:

$$\forall x, (\forall a, \, \nu_a(x) > \tau^{\text{void}}) \Rightarrow \exists \alpha \text{ such that } x \in U_\alpha \text{ and } T_\alpha \in \text{Void} \tag{14}$$

Partial membership of particles is allowed in zones. Note that (12-14) are declarative rather than operational. If necessary for modelling, the Isle label in (13) can be understood to override any preexisting Unit labels for the clustered proteins such that each protein remains associated with a single protein-containing zone as required by (12).

One choice for $\mu_a$ and $\nu_a$ is to use radial distance functions from protein centres or edges. This approach has already been employed in continuum-based attempts to identify lipid fingerprints[26]. The main advantages of this are simplicity, minimal modelling bias, and ease of implementation. Increasing the fingerprint radius can capture more loosely associated lipids, though also risks including irrelevant molecules. Thus, while radial fingerprints provide a reasonable baseline, alternative constructions may be preferable for certain applications. For concreteness and ease of visualization below, fingerprints may be interpreted using a radial construction, although no specific functional form is required by the framework. For particle membership $\theta_a$, a natural choice is to construct this from quantum-derived density fields or solvent-accessible surfaces[27,28], providing a physically grounded representation of particle extent and playing a role analogous to fingerprint scores on the zone scale. The extent to which different choices of thresholds and membership functions influence quantitative outcomes is a natural topic for future investigation.

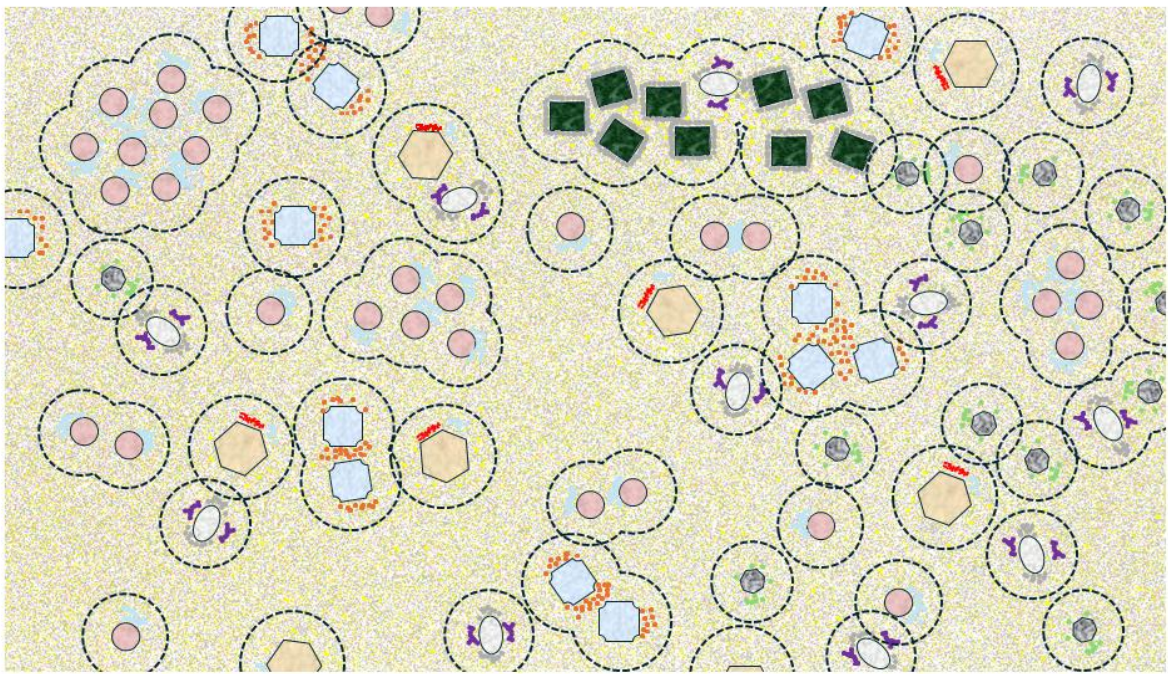

**Figure 1.** Schematic depiction of membrane zonation on a topological space. The illustrated cover is one of many possible choices and is based on radial lipid fingerprint membership from protein centres. Distances from protein edges is another viable choice. Boundaries of Void zones are omitted for clarity but are assumed to overlap with surrounding lipid fingerprints. For realistic lipid density patterns, see time-averaged molecular dynamics simulations[4].

# Hierarchical Zone Attributes

With a topological definition of zonation in place, we now introduce a hierarchical scheme for assigning data to zones. Previously, it was explained that zones are characterized by primary (1°), secondary (2°), tertiary (3°), and quaternary (4°) attributes, akin to the structure hierarchy for proteins and nucleic acids[1]. We will define the zone attribute hierarchy more broadly for convenient use with mathematical structures like hypergraphs and sheaves. For each zone $\alpha$, we associate a poset of attribute collections $(\{1_\alpha^\circ, 2_\alpha^\circ, 3_\alpha^\circ, 4_\alpha^\circ\}, \leq)$ with the order relation:

$$\mathcal{P}_\alpha = (1_\alpha^\circ \leq 2_\alpha^\circ \leq 3_\alpha^\circ \leq 4_\alpha^\circ), \tag{15}$$

where $1_\alpha^\circ$ is the set of particles in zone $\alpha$, or congruently, the attributes that are internal to particles, $2_\alpha^\circ$ are attributes associated with particle interactions in zone $\alpha$, $3_\alpha^\circ$ are attributes internal to zone $\alpha$, and $4_\alpha^\circ$ are interactions of zone $\alpha$ with other zones. Thus, 1° and 3° are intrinsic attributes while 2° and 4° are relational attributes on the particle and zone scales, respectively.

To elaborate on the meaning of "intrinsic" and "relational," this scheme can be understood through attributed hypergraphs (Figure 2A). Intrinsic attributes are assigned to nodes, while relational attributes are assigned to hyperedges. Each hyperedge carries two types of data: inherited components derived from the intrinsic attributes of its incident nodes, and nascent components that are not present in any individual node. For a system with a single isolated node, intrinsic attributes must still be represented in the relational layer to maintain consistency. This is achieved through a family of self-loop hyperedges, each carrying a portion of the node's intrinsic attributes, such that their aggregation (via direct sum, union, or other appropriate operation) reconstructs the full intrinsic attribute set. In this case, there are no nascent components; the relational layer purely reflects the intrinsic structure. When a hyperedge connects two or more nodes, it inherits components from each attached node and additionally carries nascent relational attributes. The inherited components from all hyperedges incident to a given node should aggregate to recover that node's intrinsic attributes exactly, ensuring consistency between the intrinsic and relational layers.

There are numerous possibilities for the exact meaning of the binary relation $\leq$ in $\mathcal{P}_\alpha$. It may be used to express the idea that attribute levels "build on" previous levels. The aggregation of relational attributes of zone $\alpha$ may be understood as never "less than" the aggregation of intrinsic attributes of $\alpha$ in the sense that intrinsic attributes are always fully represented in the inherited relational layer. In the hypothetical case that the system contains only a single particle, the hierarchy collapses: 1° = 2° = 3° = 4°. This is because particle relational attributes are properties of the particle itself, the zone only contains a single particle, so zone-level attributes equal particle attributes, and there is only one zone, so zone relational attributes match zone intrinsic attributes. Importantly, attributes are not ordinarily homogeneous across particles and zones. For example, there may be different curvature profiles at different locations[4]. This heterogeneity has important implications for particle and zone interactions. Naturally, a zone is merely a particle on a higher scale, and this scheme can be applied iteratively at increasingly larger and smaller scales[29].

This intuition carries over naturally to the topological formulation (Figure 2B). Replacing nodes with open sets $U_i$ and hyperedges with overlap regions, an attribute-valued presheaf or sheaf assigns intrinsic and relational attributes to the former and latter, respectively. Overlap regions play the role of interaction interfaces: intrinsic attributes from the participating regions are restricted to the overlap, forming inherited relational components that encode how intrinsic properties are presented for interaction. Every open set participates in at least the self-overlap $U_i \cap U_i = U_i$, which supports inherited relational components corresponding to restrictions of intrinsic attributes, ensuring that intrinsic attributes are always supported in the relational layer. Additional overlaps with neighbouring regions introduce further interaction interfaces and carry nontrivial nascent relational attributes. As a result, intrinsic attributes are always embedded in a surrounding network of overlaps that provides the necessary context for interaction.

For pedagogical clarity or practical shorthand, attributes may be expressed in terms of discretized or geometric representatives, such as identifying primary structure with a set of particles and secondary structure with their spatial arrangement[1]. The poset $\mathcal{P}$ is a conceptual organizing principle rather than an empirically fitted object and is intended to be broadly applicable.

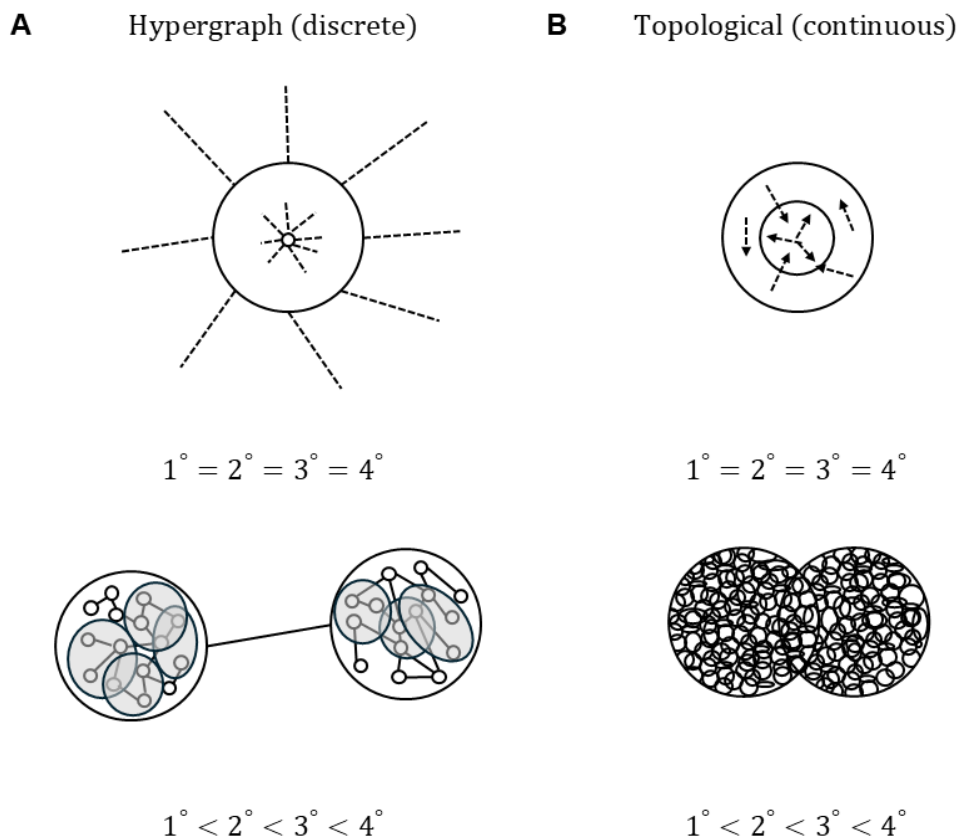

**Figure 2.** Instances of the hierarchical attribute scheme with hypergraph (**A**) and topological (**B**) systems.

## Interactions and Multiscale Structure

As discussed, interactions between particles and zones will be formalized through their intersections. Restriction maps determine the intrinsic contribution to interactions on overlaps, while the full interaction consists of this inherited part together with a nascent relational component that arises only at the intersection. To account for poses, I will modify a presheaf $\mathcal{F}$ on the category of open sets of $X$ into a pose-aware presheaf $\mathcal{F}^*$, using the superscript * to indicate pose-awareness. For attribute choices where restriction and gluing are physically meaningful, the pose-aware restriction maps may be required to satisfy the sheaf axioms, yielding a pose-aware sheaf as a special case. Throughout, we allow $\mathcal{F}^*$ to take values in an abelian category[11,30]. The underlying base space is $X$; poses act through auxiliary operators on stalks and restriction maps, including the selection or parametrization of the attributes assigned to a given open set.

Pose-awareness is introduced because the contribution of intrinsic attributes to interactions depends on more than mere set inclusion. Even when an open set matches the geometric shape of a particle, it specifies only which points belong to the region and does not come equipped with a choice of relative coordinate frame, surface normal direction, or other interaction-relevant geometric structure. Encoding such embedding information as attribute assignments would conflate intrinsic properties with interaction-specific context, erasing the distinction between what an object is and how it interacts. This separation is also practical: multiple copies of the same object may share identical pose-unaware intrinsic attributes while differing slightly in pose, allowing these differences to be accounted for through small, local transformations.

For each structured open set $U_i^* = (U_i, Y_i, T_i)$, the stalk $\mathcal{F}(U_i)$ or $\mathcal{F}^*(U_i)$ represents intrinsic attributes supported on $U_i$. For any nonempty intersection $U_{i_0} \cap \dots \cap U_{i_k} \neq \emptyset$ the stalk $\mathcal{F}(U_I)$ or $\mathcal{F}^*(U_I)$ represents relational attributes supported on interactions. We now define a pose-aware pre-restriction map:

$$S_{i_j}^* = S_{i_j}^*(Y_{i_j}) : \mathcal{F}(U_{i_j}) \to \mathcal{F}^*(U_{i_j}), \tag{16}$$

This can be composed with the ordinary restriction map $\rho_{i_j \to I}$ to create a new pose-aware restriction map:

$$\rho_{i_j \to I}^* \equiv \rho_{i_j \to I} \circ S_{i_j}^* : \mathcal{F}(U_{i_j}) \to \mathcal{F}^*(U_I). \tag{17}$$

Since there will be multiple $U_{i_j}^*$ contributing, these contributions can be directly summed:

$$\iota_I^* = \bigoplus_{j=0}^{k} \rho_{i_j \to I}^* : \bigoplus_{j=0}^{k} \mathcal{F}^*(U_{i_j}) \to \mathcal{F}^*(U_I). \tag{18}$$

Let $\text{im}(\iota_I^*) \subseteq \mathcal{F}^*(U_I)$ and $\text{coker}(\iota_I^*) \cong \mathcal{F}^*(U_I)/\text{im}(\iota_I^*)$[11]. There is a short exact sequence:

$$0 \to \text{im}(\iota_I^*) \to \mathcal{F}^*(U_I) \to \text{coker}(\iota_I^*) \to 0, \tag{19}$$

where $\text{coker}(\iota_I^*)$ represent nascent data that is not inferred from the data on individual $U_{i_j}$ but characterizes interactions. The short exact sequence expresses that all data on the overlap $U_I$ consist of two parts: the image of pose-aware restrictions from the participating regions (intrinsic contributions) and a residual quotient (nascent relational data) representing new structure that arises only on the overlap (Figure 3).

In overlaps involving multiple open sets, restriction maps must be compatible so that successive compositions agree on higher intersections. For nested overlaps $J \subset I$, functoriality requires:

$$\rho^*_{I \to J} \circ \rho^*_{i_j \to I} = \rho^*_{i_j \to J}, \tag{20}$$

The pose operators $S^*_{i_j}$ are assumed to be compatible with restriction. If a splitting is chosen (or guaranteed), then

$$\mathcal{F}^*(U_I) \cong \mathrm{im}(\iota^*_I) \oplus \mathrm{coker}(\iota^*_I). \tag{21}$$

The map $\iota^*_I$ encodes a specific choice of how intrinsic attributes are taken to contribute to overlap regions, and different admissible choices lead to different decompositions of $\mathcal{F}^*(U_I)$ into inherited and residual components. Importantly, these choices are not arbitrary: experimental or simulation data can be used to constrain and refine the restriction maps $\rho^*_{i_j \to I}$ or the aggregate restriction maps $\iota^*_I$ by identifying which intrinsic attributes consistently contribute to interaction behaviour on overlaps. Attributes that are repeatedly observed to participate in interactions may be incorporated into $\mathrm{im}(\iota^*_I)$, while remaining information is captured in $\mathrm{coker}(\iota^*_I)$. Physical constraints therefore enter through comparison, consistency, and empirical grounding.

There is considerable flexibility in the choice of attributes, provided these compatibility conditions are satisfied. A sufficient structural assumption is that attribute spaces are finite-dimensional vector spaces and that the pose-aware restriction maps $\rho^*_{i_j \to I}$ are linear and satisfy the usual cocycle condition. In realistic models, intrinsic attributes may be scalar, vector, or tensor fields representing quantities such as electrostatic charge or hydrophobicity. Relational attributes may then include nascent properties that arise only at interfaces, such as binding affinities, mismatch or frustration measures, or coupling parameters between neighbouring regions, and are represented in $\mathrm{coker}(\iota^*_I)$. In all cases, the appearance of relational attributes can be learnable[17]. Extensions to nonlinear morphisms $\rho^*_{i_j \to I}$ are beyond the present scope.

Not all biologically relevant attributes admit meaningful restriction to arbitrary subregions. For example, attributes such as amino acid sequence may be naturally defined only on entire particles or zones and need not satisfy sheaf gluing conditions. Such attributes are treated at the presheaf level within the present framework. The sheaf axioms are imposed only for those attribute choices, such as fields or densities, for which locality and gluing have a clear physical interpretation.

For these sheaf-valued attributes, a key advantage of the formulation is that it provides a natural mechanism for gluing attributes from the particle scale to the zone scale and restricting attributes from the zone scale back to the particle scale. Define two pose-aware sheaves on $X \subset \mathbb{R}^d$, $\mathcal{F}^{*P}$ and $\mathcal{F}^{*Z}$, for particle and zone scales, respectively, corresponding to attribute choices for which locality and gluing are physically meaningful. When restricted to zone $\alpha$, that is, to attributes supported on points in $U_\alpha$, they take the form:

$$\mathcal{F}^{*P}|_{U_\alpha} = 1^\circ_\alpha \oplus 2^\circ_\alpha, \tag{22}$$
$$\mathcal{F}^{*Z}|_{U_\alpha} = 3^\circ_\alpha \oplus 4^\circ_\alpha. \tag{23}$$

The components $1^\circ_\alpha$ and $3^\circ_\alpha$ represent intrinsic attributes supported on non-overlap regions fully contained within the boundary of $\alpha$ at the particle and zone scales, respectively. The interpretation of $2_\alpha^\circ$ and $4_\alpha^\circ$ requires more care. These components collect relational attributes whose spatial support lies in $U_\alpha$, but whose contributing particles or zones need not be entirely contained within $\alpha$. At the particle scale, $2_\alpha^\circ$ includes interaction data arising from particle overlaps supported within $U_\alpha$, even when some of the participating particles extend beyond the boundary of $\alpha$. Similarly, at the zone scale, $4_\alpha^\circ$ includes overlap data between $\alpha$ and other zones, reflecting interactions whose support lies in $U_\alpha$ but whose full specification may depend on neighbouring zones. Thus, restricting to $\alpha$ does not imply ignoring contributions from objects outside $\alpha$; rather, it restricts attention to interactions localized within $\alpha$.

Importantly, overlap structure at the particle scale does not, in general, align with overlap structure at the zone scale. Zone-scale overlaps may therefore correspond to particle-scale relational attributes that are supported well within the interior of $\alpha$, rather than being localized near its boundary. As a result, no clean separation of intrinsic and relational attributes is preserved under a change of scale. For this reason, maps between attributes from the particle to zone scales ($m_{P \to Z}$) or vice versa ($m_{Z \to P}$) combine intrinsic and relational attributes:

$$m_{P \to Z} \equiv \mathcal{F}^{*P}|_{U_\alpha} \to \mathcal{F}^{*Z}|_{U_\alpha} : 1_\alpha^\circ \oplus 2_\alpha^\circ \to 3_\alpha^\circ \oplus 4_\alpha^\circ, \tag{24}$$
$$m_{Z \to P} \equiv \mathcal{F}^{*Z}|_{U_\alpha} \to \mathcal{F}^{*P}|_{U_\alpha} : 3_\alpha^\circ \oplus 4_\alpha^\circ \to 1_\alpha^\circ \oplus 2_\alpha^\circ, \tag{25}$$

The maps $m_{P \to Z}$ and $m_{Z \to P}$ are not morphisms of the underlying sheaf, as they do not commute with restriction maps. They may represent aggregating, projecting, or otherwise reparametrizing attribute data. The map $m_{Z \to P}$ is generally lossy in the sense that multiple particle-scale configurations may correspond to the same zone-level description. Overall, this framework explains how attribute assignments remain compatible with direct-sum decompositions both within and across scales.

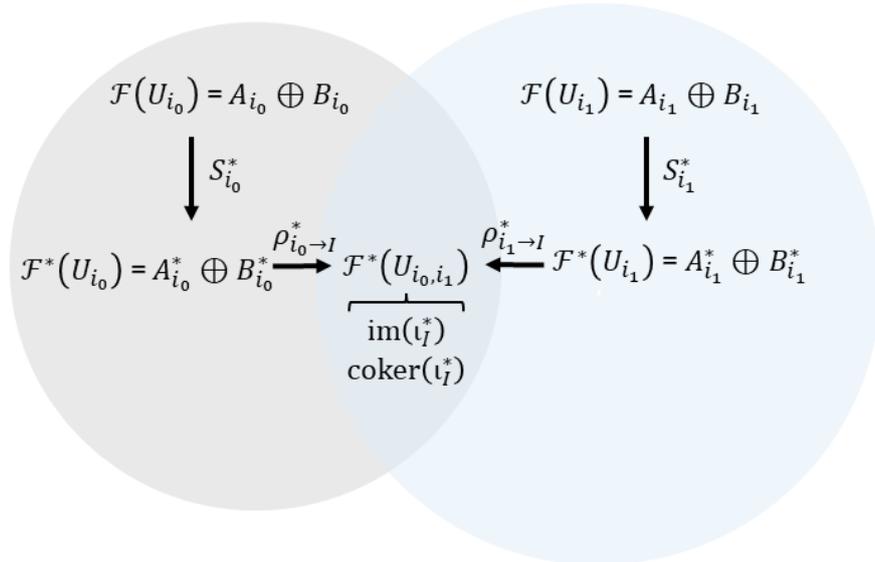

**Figure 3.** Pose-aware restriction and interaction on overlaps. Intrinsic attribute data on $U_{i_0}$ and $U_{i_1}$ are modified by pose-dependent operators $S_{i_j}^*$ before being restricted to the overlap $U_{i_0, i_1}$. The overlap stalk $\mathcal{F}^*(U_{i_0, i_1})$ contains both inherited contributions from the participating regions, given by $\mathrm{im}(\iota_I^*)$, and nascent relational attributes represented by $\mathrm{coker}(\iota_I^*)$.

## Dynamics and Energetics

The preceding sections establish a topological and categorical framework for describing particles, zones, and their interactions. A natural next step is to make this framework compatible with dynamical formalisms from physics such as Hamiltonian mechanics[31]. The goal of this section is exploratory rather than prescriptive. Rather than fixing a unique dynamical model, we suggest some constructions that could be used to endow the framework with physical realism. The discussion below should be read as illustrating how Hamiltonian mechanics might be layered onto the sheaf-theoretic structure, not as asserting that it must be done in this way. Indeed, expressions introduced here are not intended as explicit physical models, but as schematics for a broader class of admissible energy functionals. Furthermore, the assumption that attribute assignments take values in an abelian category is not meant to imply that physical configurations themselves are linear or additive. Rather, it reflects the level at which attributes are organized, decomposed, and compared, prior to their realization as dynamical variables in a Hamiltonian phase space.

Given the multiscale nature of the framework, it is natural to consider related but distinct dynamical descriptions at the particle and zone scales. One way to formalize this is to consider two time-parameterized families of pose-aware sheaves, $\{\mathcal{F}_t^{*P}\}_{t\in\mathbb{R}}$ and $\{\mathcal{F}_t^{*Z}\}_{t\in\mathbb{R}}$, defined on the fixed topological space $X \subset \mathbb{R}^d$, representing particle- and zone-scale descriptions, respectively[32]. At each fixed time $t$, $F_t^{*P}$ and $F_t^{*Z}$ may be viewed as ordinary pose-aware sheaves on $X$; time dependence enters through the evolution of the data assigned to open sets, not through changes to the base space. It is natural to endow the stalks of these sheaves with additional geometric structure, provided it is consistent with assumptions in the previous section, such as that attribute spaces take values in an abelian category[11,30]. In particular, one may further assume that suitable geometric realizations of these attribute spaces (or moduli spaces constructed from them) admit symplectic or Poisson structures, and that pose-aware restriction maps $\rho_{i_j\to I}^*$ are Poisson maps[33,34]. For example, the moduli spaces of the sheaves can inherit Poisson structures through Ext-pairings and trace maps[35-37].

Each attribute level $k° \in \{1°, 2°, 3°, 4°\}$ may then be associated with a collection of degrees of freedom. To avoid overcounting, it may be sensible to restrict relational degrees of freedom to nascent components; those not already determined by intrinsic assignments. One way to formalize this is to work with $\text{coker}(\iota_I^*)$ and ignore $\text{im}(\iota_I^*)$ when converting attributes to energy[11]. Alternatively, one may assume that each spatial point $x$ carries its own energy contribution computed locally from the stalk.

When a canonical description is convenient, degrees of freedom may be written schematically as coordinate-momentum pairs, $k° = (q_n^{k°}, p_n^{k°})$, with analogous constructions $(q_n^{Y_P}, p_n^{Y_P})$ and $(q_n^{Y_Z}, p_n^{Y_Z})$ for pose variables if required[31]. Here, the index $n$ labels components of an attribute or pose space. Under this interpretation, the total set of degrees of freedom is fixed; changes in zonation correspond to reweighting, recoupling, or redistribution of existing variables, not to their creation or destruction[38]. Although attributes may already be pose-aware, the pose variables themselves may be treated as independent degrees of freedom by modeling choice.

To allow structured open sets to evolve continuously while maintaining membership values in $[0,1]$, membership functions $\theta_i$ may be reparameterized through smooth sigmoid maps and be included in $q_n^{Y_P}$ and $q_n^{Y_Z}$. Depending on the modelling context, attributes may be treated either as discrete variables (leading to finite-dimensional phase spaces) or as spatially distributed fields over $X$ (leading to field-theoretic Poisson brackets)[33]. These choices reflect different levels of resolution rather than different theoretical commitments, and hybrid constructions are also conceivable.

There are many reasonable choices for the specific degrees of freedom associated with each attribute level. At the primary level, $q^{1°}$ may include quantities such as bond lengths and bond angles, with conjugate momenta representing vibrational or rotational motion. At the secondary level, $q^{2°}$ may encode interaction descriptors such as binding affinities or contact propensities, with conjugate momenta representing interaction fluxes, sensitivities, or other response variables associated with changes in those interactions. At the tertiary level, $q^{3°}$ may describe collective zone-internal properties such as curvature, thickness, or undulation[4], with corresponding deformation momenta. Coarse thermodynamic variables such as temperature or pressure may also be included to represent unresolved particle-scale motion. At the quaternary level, $q^{4°}$ may represent inter-zone relational quantities such as curvature mismatch or chemical potential differences; when treated dynamically, these collective mismatch variables may be supplemented with conjugate momenta representing their effective rates of change. In addition to membership functions, pose variables $q^{Y^P}$ and $q^{Y^Z}$ may include orientations and geometric parameters. Together with their conjugate momenta, they could allow dynamic evolution of structured open sets and smooth handling of zone formation, deformation, and disappearance[39,40].

To obtain dynamics, one must supply energy functionals. A natural organizational principle is to decompose energy into intrinsic versus relational and potential ($V$) versus kinetic ($T$) contributions:

$$H^P = V^{1°} + T^{1°} + V^{2°} + T^{2°}, \qquad (26)$$
$$H^Z = V^{3°} + T^{3°} + V^{4°} + T^{4°}. \qquad (27)$$

Each energy term $V^{k°}$ or $T^{k°}$ can be understood as a functional defined on the corresponding stalks and pose variables. Here, intrinsic energies may be computed as valuations over the nerve of the cover[41], ensuring that each spatial point contributes exactly once, while relational energies may be summed over nonempty overlaps and interpreted as corrective interaction terms rather than additional spatial contributions. This separation can be imposed to avoid energetic redundancy between intrinsic and relational attributes on overlapping regions. If necessary, weighting schemes such as partitions of unity or normalized overlap weights may be introduced to ensure invariance under cover refinement and stability under reparameterization[42-44].

This decomposition is not unique. Different modelling choices such as how pose variables enter energy terms or how relational energies are weighted can lead to qualitatively different dynamics[45-47]. For example, computing a curvature energy $V^{3°}$ may require both a curvature attribute $q^{3°}$ and information about its spatial extent encoded in $q^{Y^Z}$. The appropriate level of detail depends on the intended resolution of the model.

Given a Hamiltonian and a Poisson structure, one formally obtains equations of motion via Hamilton's equations[48]. To represent non-equilibrium behaviour typical of biological membranes, it may be reasonable to supplement these with phenomenological dissipation terms such as a Rayleigh dissipation potential[49]. In principle, such dissipation potentials could be constructed or parametrized from intrinsic and relational attributes.

Taken together, these considerations suggest that the pose-aware sheaf framework admits Hamiltonian-style formulations. At the same time, the framework is deliberately permissive: many choices of degrees of freedom, Poisson structures, energies, and scale-bridging maps are possible, and only a subset will yield stable, interpretable, or biologically relevant behaviour. Identifying those subsets and understanding their consequences remains an open problem and a natural direction for future work.

## Extensions

It is useful to distinguish between instantaneous zones and time-extended zones, which may be called "zone ensembles". A zone ensemble can be defined as all the states a zone passes through while performing a function. Ensemble averages make it easier to visualize features such as lipid fingerprints by highlighting components with long residency times[4]. However, because each realization fluctuates, different ensembles will not be identical. It can therefore be useful to define an "ensemble of averages", where each time point represents a statistical median or mean, or a "representative ensemble" capturing typical dynamics without being an average. A key challenge for membrane biology will be to connect average or representative structures to the dynamic, non-equilibrium states observed in living cells. Indeed, there is likely substantial variability even within the populations of well-studied zones that have so far been treated primarily as single entities, such as caveolae and clathrin-coated pits[50,51].

A practical route forward is to compare zone states and ensembles across multiple contexts that approximate in vivo conditions. Combining data from structural biology methods such as cryo-electron microscopy and molecular dynamics simulations may help build representative ensembles from resolved states. MemProtMD is closest to achieving an integrated pipeline for resolving lipid fingerprints with molecular dynamics, but currently only uses dipalmitoylphosphatidylcholine[52,53], while the true membrane is a giant proteolipid code[1]. A long-term objective would be the creation of a dynamic database of zones, ensembles, and attributes, integrating experimental and computational data through reinforcement learning or other data-driven paradigms. Such a database could serve as a living repository for the proteolipid code.

The framework developed here represents one possible formalization of the proteolipid code but can be generalized through alternative definitions or mathematical structures. For greater granularity, each particle could be assigned its own fingerprint relative to all other components, so that the open cover at the zone scale is centered on each individual particle rather than proteins alone. This would enable explicit tracking of lipids and other small molecules, albeit likely at higher computational cost. Additionally, it should be noted that, in a similar way that each protein has a lipid fingerprint[4], each zone should also possess a "fingerprint" of other zones, which might be used to make predictions about zone adjacency. More speculative extensions might incorporate relativistic or quantum effects, for example, assigning zones their own proper time or allowing tunneling between configurations. Thus, the heuristic presented here could, in principle, extend to systems ranging from membranes to astrophysical or subatomic scales.

A particularly natural generalization involves topos theory[32], which provides an internal logical structure that extends the concept of a topological space. In this formulation, the membrane is represented by a topos $T$, where open sets are replaced by region objects and points by epimorphisms $U_i^* \to 1_T$. Membership in particle and zone objects is defined internally through characteristic morphisms $\theta_i : X \to [0,1]$, which specify graded inclusion thresholds $\tau_P, \tau_Z \in (0,1)$. Interaction morphisms $U_{i_0}^* \times U_{i_1}^* \to [0,1]$ could then quantify how components participate in one another's fingerprints. This topos generalization naturally accommodates fuzzy or uncertain membership while preserving categorical consistency.

# Conclusions

Here I represented membrane structure algebraically, opening doors for further mathematical development and computational applications. The concept of a zone bridges micro and macro scales through a hierarchical classification of attributes. This formalism admits Hamiltonian-style formulations consistent with its sheaf-theoretic decomposition, although I resist committing to a specific dynamical realization. A key advantage of this framework is its capacity to integrate heterogeneous data while preserving mathematical consistency and physical interpretability. For example, relational attributes explicitly record how intrinsic attributes are combined and transformed into interaction capabilities. The framework thus provides the organizational structure that membrane biology has long lacked.

While the current formulation may be sensitive to choices of membership functions, thresholds, attribute assignments, weights used in dynamical constructions, and other modelling decisions, standardized choices should become more apparent after this framework is applied to real data. This article is intended to formalize the proteolipid code mathematically and introduce relevant concepts as a foundation for future research, thus a worked application is beyond the present scope. The advantage of this framework is its alignment with both formal structure and experimentally accessible zone states, combining theory and data into one.

# Declaration on the use of Generative AI

The following generative AI applications were used during the brainstorming, writing, and editing phases of this manuscript: ChatGPT (2024/25 versions, OpenAI, San Francisco, CA; https://chatgpt.com), Claude (2025 versions, Anthropic, San Francisco, CA; https://claude.ai), and q.e.d. Science (2025 versions; https://www.qedscience.com). These tools were used intermittently during 2024-2025, corresponding to the publicly available model versions current at the time of use. This article builds on human-conceived ideas such as the proteolipid code, zonation, and zone attributes[1]. Of note, the following foundational concepts originated from T.A.K.'s reasoning without direct AI assistance: the structured open set $U_i^*$, zone thresholds, $\mathcal{P}_\alpha = (1_\alpha^\circ \leq 2_\alpha^\circ \leq 3_\alpha^\circ \leq 4_\alpha^\circ)$, zone ensembles, particle-to-zone scale bridging maps. Other content is an outcome of both human and machine reasoning. T.A.K. takes full responsibility for all content. All prompts and outputs associated with this research are retained and can be made available upon request. They are not included as publicly visible supplementary files because the prompt-output corpus is interwoven with confidential concepts and developmental material intended for subsequent manuscripts, and public release could constitute prior dissemination of material planned for future publication. Access can be arranged under confidentiality or controlled-access terms if required.

# Declaration of competing interests

No competing interests are declared by T.A.K.

# Acknowledgements

I thank Michael Overduin and Peijun Zhang for discussions.

# Funding

This work was funded by a Clarendon Fund Scholarship in partnership with a Nuffield Department of Clinical Medicine Studentship and a Magdalen College Graduate Scholarship to TAK.